\documentclass[12pt,preprint]{aastex}

\received{}
\accepted{}

\bibpunct{(}{)}{;}{a}{}{,}

\begin{document}


\author{
Michael A. Weinstein,\altaffilmark{1,2}
Gordon T. Richards,\altaffilmark{3}
Donald P. Schneider,\altaffilmark{1}
Joshua D. Younger,\altaffilmark{3}
Michael A. Strauss,\altaffilmark{3}
Patrick B. Hall,\altaffilmark{3}
Tam\'as Budav\'ari,\altaffilmark{4}
James E. Gunn,\altaffilmark{3}
Donald G. York,\altaffilmark{5,6}
and J. Brinkmann\altaffilmark{7}
}

\altaffiltext{1}{Department of Astronomy and Astrophysics, The Pennsylvania State University, 525 Davey Laboratory, University Park, PA 16802.}
\altaffiltext{2}{Department of Physics, Astronomy, and Geophysics, Connecticut College, Box 5622, 270 Mohegan Avenue, New London, CT, 06320.}
\altaffiltext{3}{Princeton University Observatory, Peyton Hall, Princeton, NJ 08544.}
\altaffiltext{4}{Department of Physics and Astronomy, The Johns Hopkins University, 3400 North Charles Street, Baltimore, MD 21218-2686.}
\altaffiltext{5}{Department of Astronomy and Astrophysics, The University of Chicago, 5640 South Ellis Avenue, Chicago, IL 60637.}
\altaffiltext{6}{Enrico Fermi Institute, The University of Chicago, 5640 South Ellis Avenue, Chicago, IL 60637.}
\altaffiltext{7}{Apache Point Observatory, P.O. Box 59, Sunspot, NM 88349.}

\title{An Empirical Algorithm for Broad-band Photometric Redshifts of Quasars from the Sloan Digital Sky Survey}

\begin{abstract}
We present an empirical algorithm for obtaining photometric redshifts
of quasars using 5-band Sloan Digital Sky Survey (SDSS) photometry.
Our algorithm generates an empirical model of the quasar
color-redshift relation, compares the colors of a quasar candidate
with this model, and calculates possible photometric redshifts.  Using
the 3814 quasars of the SDSS Early Data Release (EDR) Quasar Catalog
to generate a median color-redshift relation as a function of redshift
we find that, for this same sample, 83\% of our predicted redshifts
are correct to within $|\Delta z|<0.3$.  The algorithm also determines the
probability that the redshift is correct, allowing for even more
robust photometric redshift determination for smaller, more restricted
samples.  We apply this technique to a set of 8740 quasar candidates
selected by the final version of the SDSS quasar-selection algorithm.
The photometric redshifts assigned to non-quasars are restricted to a
few well-defined values.  In addition, 90\% of the objects with
spectra that have photometric redshifts between 0.8 and 2.2 are
quasars with accurate ($|\Delta z|<0.3$) photometric redshifts.  Many
of these quasars lie in a single region of color space; judicious
application of color-cuts can effectively select quasars with accurate
photometric redshifts from the SDSS database --- without reference to
the SDSS quasar selection algorithm.  When the SDSS is complete, this
technique will allow the determination of photometric redshifts for
$\sim 10^6$ faint SDSS quasar candidates, enabling advances in our
knowledge of the quasar luminosity function, gravitational lensing of
quasars, and correlations among quasars and between galaxies.

\end{abstract}


\section{Introduction}\label{chap:intro}

A photometric redshift (photo-$z$) is derived from the colors and
morphology of an object, rather than from its spectrum.  Since the
bandwidth of an imaging filter is typically $\sim 1000$ \AA\ while
that of a spectroscopic exposure is more like $\sim$ 1--10 \AA, it is
much faster to obtain an image than to take a spectrum, and thus
photometric redshift determination has the potential for determining
reasonably accurate redshifts for large numbers of objects with a
minimum of telescope time.  In recent years, this technique has been
applied to galaxies with great success
\markcite{Br1997,C1999,mib+04,bfb+04}(e.g., {Brunner} {et~al.} 1997; {Connolly} {et~al.} 1999; {Mobasher} {et~al.} 2004; {Ben{\'{\i}}tez} {et~al.} 2004) as a result of the strong,
discontinuous features found in the spectra of galaxies, such as the
4000 \AA\ break, which cause galaxy colors to change substantially
with redshift.

High-redshift quasars have a similar discontinuity as a result of
hydrogen absorption blueward of Ly$\alpha$ emission (i.e., the
Ly$\alpha$ forest; \markcite{lyn71}{Lynds} 1971) that allows for rough
determination of their redshifts from broad-band photometry
\markcite{cab+04}(e.g., {Cristiani} {et~al.} 2004).  Lower redshift quasars lack such a strong
discontinuity in optical bands, but structure in the color-redshift
relation caused by strong emission lines can be used to determine
accurate photometric redshifts if the errors in the photometry are
sufficiently small.

Early attempts at determining photometric redshifts for quasars are
described by \markcite{Hat2000}{Hatziminaoglou}, {Mathez}, \&  {Pell{\' o}} (2000) and \markcite{Wo2001}{Wolf} {et~al.} (2001) --- demonstrating a
success rate of roughly 50\% (within $|\Delta z| < 0.3$) for a few
dozen quasars.  The 17-filter COMBO-17 survey has had more success ---
identifying 192 $1.2\le z \le4.8$ quasars with an expected photometric
redshift accuracy of $\sigma_z=0.03$ \markcite{wwb+03}({Wolf} {et~al.} 2003).  \markcite{rws+01}{Richards} {et~al.} (2001b)
and \markcite{bcs+01}{Budav{\' a}ri} {et~al.} (2001) described two techniques for determining
photometric redshift of quasars from the Sloan Digital Sky Survey
(SDSS; \markcite{yaa+00}{York} {et~al.} 2000) imaging data.  In this paper we expand and
improve upon the empirical method that \markcite{rws+01}{Richards} {et~al.} (2001b, hereafter
RW01) applied to the SDSS photometry of 2625 quasars and
Seyferts presented in \markcite{R2001a}{Richards} {et~al.} (2001a).

In \S~2 we review the details of the SDSS data and previous photo-$z$
attempts for quasars.  Section~\ref{chap:newalg} describes our current
photometric redshift algorithm and how it differs from the original
one presented in RW01.  In \S\S~\ref{chap:EDRQtests}
and~\ref{chap:FEDRtests}, we test our algorithm with two sets of SDSS
data, each superior to that used in RW01.
Section~\ref{chap:FEDRtests} also discusses practical applications of
this approach.  Section~\ref{chap:summary} summarizes our work.

\section{Previous Attempts at Photometric Redshifts Using SDSS Imaging Data}

This work, RW01, and \markcite{bcs+01}{Budav{\' a}ri} {et~al.} (2001) determine photometric redshifts
for quasars by taking advantage of the quantity and quality of the
imaging data from the SDSS.  The SDSS obtains imaging data using a
wide-field multi-CCD camera \markcite{gcr+98}({Gunn} {et~al.} 1998) with five broad bands
($ugriz$; \markcite{fig+96}{Fukugita} {et~al.} 1996).  The photometric calibration of these
data is described by \markcite{hfs+01}{Hogg} {et~al.} (2001), \markcite{stk+02}{Smith} {et~al.} (2002), and
\markcite{slb+02}{Stoughton} {et~al.} (2002).  Throughout this work we use point-spread-function
``asinh'' magnitudes \markcite{lgs+99}({Lupton}, {Gunn}, \& {Szalay} 1999) that have been corrected for
Galactic reddening \markcite{sfd98}({Schlegel}, {Finkbeiner}, \&  {Davis} 1998).  \markcite{rfn+02}{Richards} {et~al.} (2002) present the quasar
selection algorithm, while \markcite{sfh+03}{Schneider} {et~al.} (2003) present the most recent
catalog of bona-fide SDSS quasars.  The SDSS's tiling algorithm and
astrometric accuracy are described by \markcite{blm+03}{Blanton} {et~al.} (2003) and
\markcite{pmh+03}{Pier} {et~al.} (2003), respectively.  Further details related to Data
Releases One (DR1) and Two (DR2) can be found in \markcite{aaa+03}{Abazajian} {et~al.} (2003) and
\markcite{aaa+04}{Abazajian et al.} (2004).

The 2625 quasars used in RW01 and \markcite{bcs+01}{Budav{\' a}ri} {et~al.} (2001) were not chosen with
one uniform quasar-selection algorithm (the quasar selection procedure
was under development at the time), but all of the objects had spectra
as well as SDSS photometry.  The empirical photometric redshift
algorithm applied by RW01 minimized the $\chi^2$ between the observed
and median colors of quasars as a function of redshift.  It correctly
predicted the redshifts of 55\% of the objects to within 0.1, and 70\%
to within 0.2, a significant advance compared to previous work on this
subject --- both in sample size and accuracy.

RW01 pointed out some problem areas for future work on their
algorithm.  ``Reddened quasars'' (see \markcite{R2003}{Richards} {et~al.} 2003 and
\markcite{hhs+04}{Hopkins et al.} 2004) usually had incorrect photometric redshifts.
Objects with an extended morphology also were more likely to yield
incorrect redshift predictions.  In addition, it was found that
quasars at certain specific redshifts below 2.2 had nearly identical
colors (``color-redshift degeneracies''), where the photometric
redshift algorithm would sometimes select the incorrect value.  RW01
discussed some improvements (e.g., weighting by redshift and assigning
probabilities to the photometric redshifts) that could be made to the
algorithm.  This paper describes some of those improvements.

The photometric redshift tests in RW01 were performed with quasars
whose redshifts were already spectroscopically known, but of course
the ultimate goal is to apply the technique to quasar candidates that
do not have spectra.  Two solutions to the problem of accounting for
non-quasars in the sample were suggested in RW01.  First, it might be
possible to differentiate between quasars and non-quasars using their
photometric redshifts; perhaps most non-quasars would be assigned
redshifts within a few narrow ranges that could be excluded a
posteriori.  Herein we test the accuracy of this suggestion.  Second,
it might be possible to select specific regions in color-space that
have very high efficiency in quasar selection.  We test this
suggestion for a small sample of objects from the SDSS Early Data
Release (EDR; \markcite{slb+02}{Stoughton} {et~al.} 2002) database, while \markcite{R2004}{Richards et al.} (2004)
describes a more powerful selection algorithm (in terms of
completeness and efficiency of quasar selection) using a much larger
sample from the DR1 database.

\section{The Algorithm}\label{chap:newalg}

Suppose we have two sets of quasars, designated as $\mathcal{S}$ and
$\mathcal{P}$.  Those in set $\mathcal{S}$ have spectra, so we know
each one's redshift, $z_{\mathrm{spec}}$.  We also have photometry for
these quasars, as measured by the Sloan Digital Sky Survey: five
magnitudes $(u, g, r, i, z)$, and their one-sigma uncertainties
$(\sigma_u, \sigma_g, \sigma_r, \sigma_i, \sigma_z)$.  The quasars in
the other set, $\mathcal{P}$, have no spectra, only photometry.  Our
goal is to obtain photometric redshifts $z_{\mathrm{phot}}$ for each
of the quasars in $\mathcal{P}$.

There are two primary steps to our algorithm.  First one must
construct an empirical color-redshift relation (CZR) using set
$\mathcal{S}$.  Then one uses this CZR to assign photometric redshifts
to each of the quasars in $\mathcal{P}$.  We give the details of these
steps below (using the quasars of the EDR catalog for set
$\mathcal{S}$).

\subsection{Construction of the CZR from the Quasars in $\mathcal{S}$}
\label{sec:algsteps1+2}

To begin, we sort the quasars of $\mathcal{S}$ into $N$ redshift bins.
As in RW01, we chose bins with centers at intervals of 0.05 in
redshift, and widths of 0.075 for $z_{\mathrm{spec}}<2.15$, 0.2 for
$2.15 \leq z_{\mathrm{spec}} < 2.5$, and 0.5 for $z_{\mathrm{spec}}
\geq 2.5$.  The bins overlap, and become wider for higher redshift, in
order to maintain enough quasars in each bin.

\begin{figure}[p]
\plotone{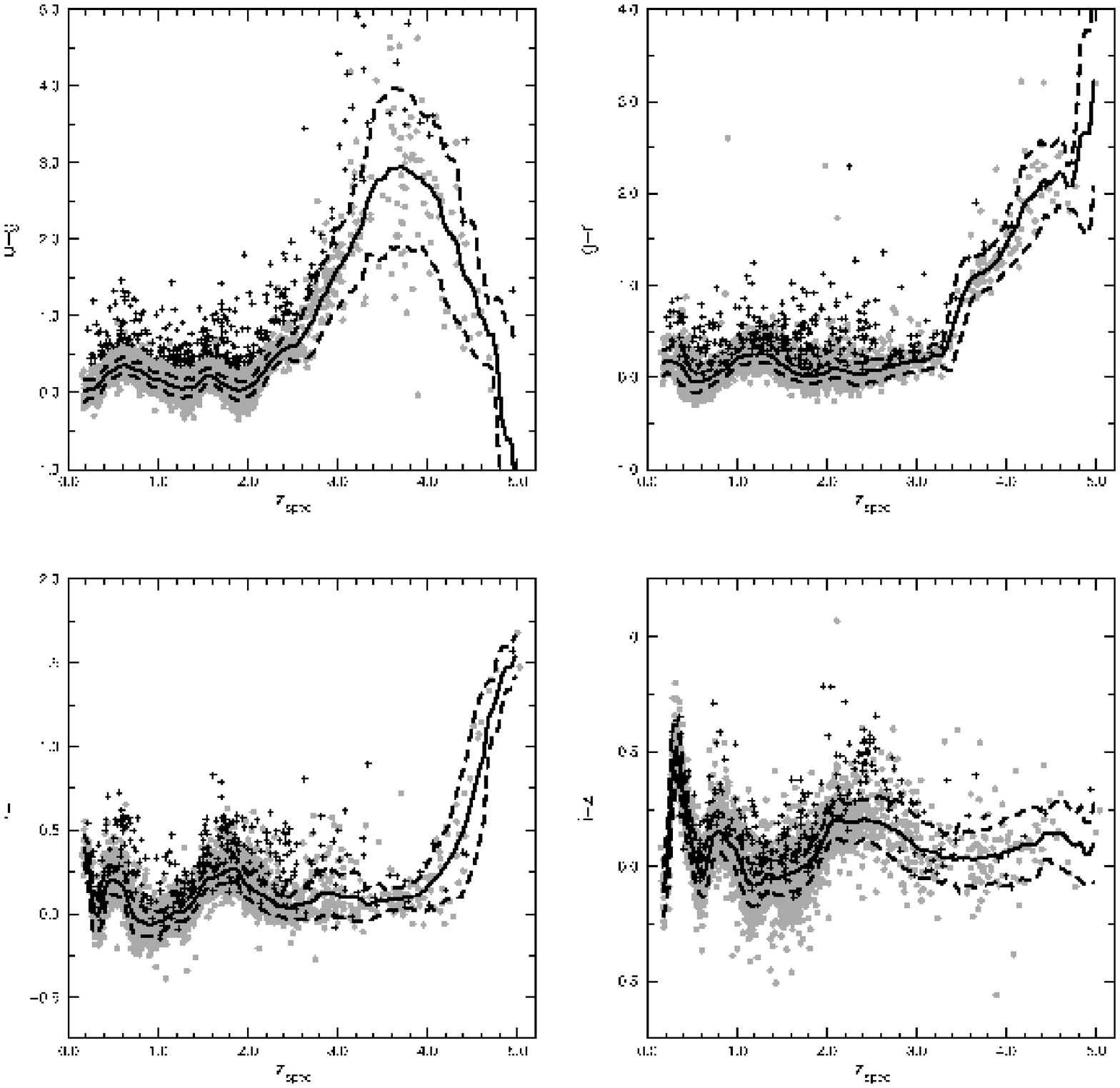}
\caption{SDSS colors vs.\ spectroscopic redshift for EDR quasars,
with CZR superimposed.  \textit{Gray points}: non-reddened quasars,
\textit{black pluses}: reddened quasars, \textit{solid line}: mean
color of CZR ($M_i^j$) for each redshift bin, \textit{dashed lines}:
1-$\sigma$ range of uncertainty in CZR colors
($M_i^j\pm\sqrt{V_i^{jj}}$).}
\label{fig:fig1}
\end{figure}

In Figure~\ref{fig:fig1}, we plot the SDSS colors of the EDR quasars
as functions of spectroscopic redshift.  Most quasars in a particular
redshift bin have very similar colors, and it is this color-redshift
relation that we wish to parameterize.  However, there are a small
fraction of quasars --- plotted with black pluses --- whose colors are
significantly redder than they should be (compared to other quasars at
the same redshift), especially in $u-g$ and $g-r$.  We call these
outliers to the CZR ``reddened quasars'' since they appear to be
reddened by internal dust extinction \markcite{R2003}({Richards} {et~al.} 2003).  To prevent these
quasars from skewing the CZR to the red, we exclude them from
$\mathcal{S}$ before constructing the CZR.

\subsubsection{Identification of the Reddened Quasars in $\mathcal{S}$}
\label{sec:algstep1}

In this work, we defined reddened quasars using $u-g$ and $g-r$.  Our
reasoning was as follows: Figure~\ref{fig:fig1} shows that $u-g$
colors can be significantly reddened.  However, the uncertainties in
$u-g$ are often quite large, due to the large uncertainties in SDSS
$u$ magnitudes.  Thus, we set a further constraint that reddened
quasars should also be fairly red in $g-r$, the color that is
second-most affected by the reddening.  If a quasar's $u-g$ color is
redder than 97.5\% of the other quasars in the same redshift bin, and
its $g-r$ color is redder than the median $g-r$ color for its redshift
bin, then we consider it to be a reddened quasar, and exclude it from
set $\mathcal{S}$.

Recently, \markcite{R2003}{Richards} {et~al.} (2003) have shown that a more effective method for
isolating reddened quasars is to make a cut on $\Delta(g-i)$: the
$g-i$ color relative to the mean $g-i$ color at that redshift.  In
addition it may not be necessary to exclude reddened quasars at all if
modal colors \markcite{hhs+04}({Hopkins et al.} 2004) are used to determine the CZR, since these
colors should be unaffected by reddened outliers.  We may alter our
algorithm in the future to take these developments into account, but
our definition of reddened quasars suffices for our purposes.

\subsubsection{Parameterizing the CZR}
\label{sec:algstep2}

Now that we have excluded the most heavily reddened quasars, we use
the remaining quasars in set $\mathcal{S}$ to construct the CZR.  In
what follows, the index $i=1,\ldots,N$ will refer to the redshift
bins.  Let $Q_i$ be the number of non-reddened quasars in the
$i^{\mathrm{th}}$ redshift bin; then the index $q=1,\ldots,Q_i$ ranges
over the non-reddened quasars in that bin.

Using the colors of the quasars in the $i^{\mathrm{th}}$ bin, we 
calculate the
mean color vector $\mathbf{M}_i$ and the color covariance matrix
$\mathbf{V}_i$ for that bin.  The four components of $\mathbf{M}_i$ are:
\[
M_i^j = \frac{1}{Q_i} \sum_{q=1}^{Q_i} x_{j,q} \hspace{1cm} (j=1,2,3,4)
\]
and the sixteen components of $\mathbf{V}_i$ are:
\[
V_i^{jk} = \frac{1}{Q_i-1} \sum_{q=1}^{Q_i} (x_{j,q}-M_i^j)(x_{k,q}-M_i^k) \hspace{1cm} 
           (j,k=1,2,3,4),
\]
where $j$ and $k$ represent colors, and $x_{1,q}$, $x_{2,q}$,
$x_{3,q}$, and $x_{4,q}$ are the $u-g$, $g-r$, $r-i$, and $i-z$ colors
(respectively) of the $q^{\mathrm{th}}$ non-reddened quasar in the
$i^{\mathrm{th}}$ redshift bin.

We then refine $\mathbf{M}_i$ and $\mathbf{V}_i$ by utilizing the
following iterative procedure: (1) assume that the color distribution
of the non-reddened quasars in the $i^{\mathrm{th}}$ bin is a
four-dimensional multivariate normal distribution with mean
$\mathbf{M}_i$ and covariance $\mathbf{V}_i$, (2) throw out any
quasars in that bin with colors that lie in the outermost one percent
of this hypothetical distribution, and (3) recalculate $\mathbf{M}_i$
and $\mathbf{V}_i$ for the remaining quasars in the bin.  Repeat this
procedure until $\mathbf{M}_i$ and $\mathbf{V}_i$ are unaltered; for
the EDR data, we found that two iterations (removing roughly 7\% of the
non-reddened quasars) were sufficient.

The final $\mathbf{M}_i$ and $\mathbf{V}_i$, for all $i$, define the
empirical color-redshift relation, or CZR.  In Figure~\ref{fig:fig1},
we plot a representation of the CZR constructed from the non-reddened
quasars in the EDR quasar catalog.  In the four plots ($j=1,2,3,4$),
$M_i^j$ (\textit{solid lines}) and $M_i^j \pm \sqrt{V_i^{jj}}$
(\textit{dashed lines}) are plotted versus redshift.

\begin{figure}[p]
\plotone{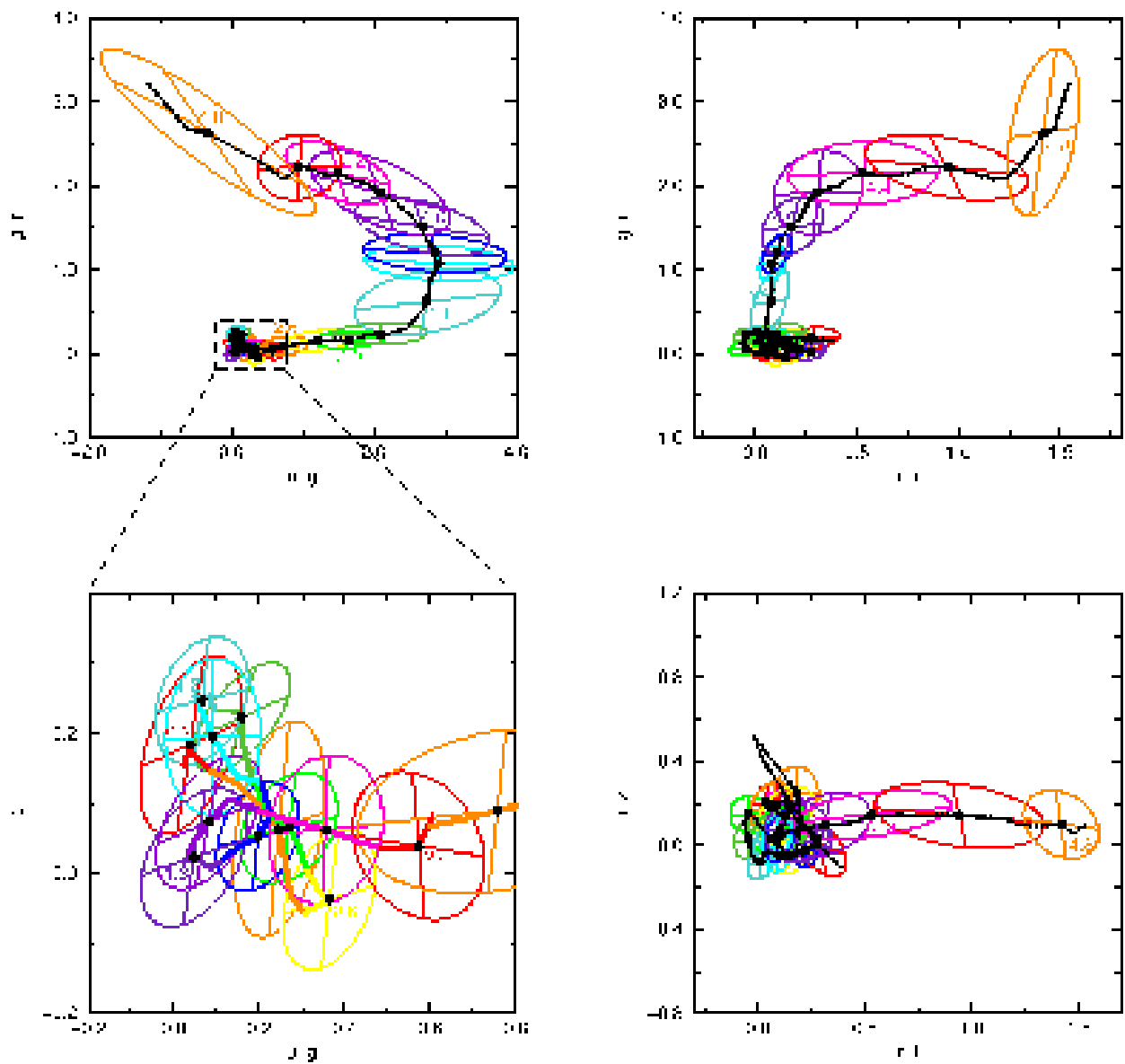}
\caption{Color-space plot of CZR, as determined by the EDR quasars.
\textit{Top-left, top-right, and bottom-right plots} are
two-dimensional projections of color-space --- \textit{black dotted
line}: CZR track, \textit{colored ellipses}: 1-$\sigma$ error
ellipsoids of CZR for selected redshift bins, \textit{numbered points}
mark selected redshifts along the CZR.  \textit{Bottom-left plot} is
an expansion of the indicated region in the \textit{top-left plot};
the CZR track in this plot is colored to match the error
ellipsoids. The dots are at intervals of 0.2 in
redshift.}
\label{fig:fig2}
\end{figure}

Another way to visualize the CZR is as a track in color-space, made of
a series of four-dimensional multivariate Gaussian distributions (one
for each redshift bin).  Each Gaussian has an ellipsoidal
cross-section.  In Figure~\ref{fig:fig2}, we plot the EDR quasar CZR
in two-dimensional projections of color-space.  The CZR track for
$z<2.2$ is completely contained in one small region of color-space
$\sim 0.3$ magnitudes across.  Quasars at these redshifts have similar
colors, since the optical/UV continuum of a quasar spectrum is
well-approximated by a power-law (which is invariant with redshift).
This illustrates why it was so difficult to obtain photometric
redshifts for these quasars using photographic plate photometry, with
errors of $\sim 0.1$ magnitudes.  

Now that we have constructed the CZR, we are ready to determine photometric 
redshifts.

\subsection{Obtaining Photometric Redshifts for Quasars in 
$\mathcal{P}$}\label{sec:algstep3}

For a particular quasar in $\mathcal{P}$ for which we want to obtain a
photometric redshift $z_{\mathrm{phot}}$, let the vector
$\mathbf{X_0}$ be its observed colors $(u-g, g-r, r-i, i-z)$, and let
the matrix $\mathbf{V_0}$ be the covariance matrix of its colors.  The
latter can be derived from the one-sigma uncertainties in its observed
magnitudes $(\sigma_u, \sigma_g, \sigma_r, \sigma_i, \sigma_z)$,
assuming that the errors are uncorrelated with each
other\footnote{Using repeat scans of SDSS stripe 82, it has been shown
that the errors in the magnitudes correspond to the observed scatter
and are minimally correlated (R. Scranton, private communication).}:
\[
\begin{array}{ccc}
\mathbf{V_0} & = & \left( \begin{array}{cccc}
		   \sigma_u^2+\sigma_g^2 & -\sigma_g^2 & 0 & 0 \\
		   -\sigma_g^2 & \sigma_g^2+\sigma_r^2 & -\sigma_r^2 & 0 \\
		   0 & -\sigma_r^2 & \sigma_r^2+\sigma_i^2 & -\sigma_i^2 \\
		   0 & 0 & -\sigma_i^2 & \sigma_i^2+\sigma_z^2
		   \end{array} \right)
\end{array}
\]

We now compute the chi-square value, $\chi^2_i$, between the observed
colors of the quasar, and the CZR's $i^{\mathrm{th}}$ redshift bin:
\[
\chi^2_i = (\mathbf{X_0}-\mathbf{M}_i)^T (\mathbf{V_0}+\mathbf{V}_i)^{-1} 
	   (\mathbf{X_0}-\mathbf{M}_i) \hspace{1 cm} (i=1,\ldots,N)
\]
and from this $\chi^2$ value, derive the probability $P_i$ that the
quasar's redshift lies in the $i^{\mathrm{th}}$ redshift bin:
\begin{eqnarray*}
P'_i & = & 
   \frac{W_i \exp(-\chi^2_i/2)}{4\pi^2 |\mathbf{V_0}+\mathbf{V}_i|^{1/2}}\\
P_i & = & \frac{P'_i}{\sum_{i=1}^N P'_i} \hspace{2 cm} (i=1,\ldots,N),
\end{eqnarray*}
where $P_i$ is normalized so that its sum over all $i$ is unity.

$W_i$ is the a priori probability that a quasar has a redshift in the
$i^{\mathrm{th}}$ redshift bin.  If the redshift distributions of
$\mathcal{S}$ and $\mathcal{P}$ are expected to be similar, then $W_i$
is the fraction of quasars in $\mathcal{S}$ (both reddened and
non-reddened) that lie in the $i^{\mathrm{th}}$ redshift bin;
otherwise $W_i$ should be equal to $1/N$ for all $i$.

If the object is identified as an extended source by the SDSS
photometric pipeline, it is probably at a redshift of less than one.
In that case, we only calculate probabilities $P_i$ for the $N'$
redshift bins that lie between $z=0$ and $z=1$ (where this upper limit
is an argument to the algorithm), and $N$ is replaced with $N'$ in the
formulae for $P'_i$ and $P_i$ above.  (For an extended object, $W_i$
is handled in exactly the same way as it is for a point-source.)  Of
course there are dangers in taking this approach.  For example, at
fainter magnitudes the SDSS's star-galaxy separation can break down,
and superposition of sources can mimic extended objects.

Finally, to obtain a photometric redshift for the quasar, we search
for groups of consecutive redshift bins for which each $P_i$ exceeds
some threshold value (our choice was $1/N$).  Each such group defines:
\begin{itemize}
\item a photometric redshift (the redshift bin in the group with the
largest $P_i$), 
\item an approximate range to the photometric redshift (obtained from
the lowest and highest redshifts in the group), and
\item a probability (or ``confidence-level'') that the actual redshift
is within this range (the sum of $P_i$ for all redshift bins in the
group).
\end{itemize}
One can either restrict one's attention to the photometric redshift
with the highest probability, or list multiple choices for the
photometric redshift with their associated probabilities.

Figure~\ref{fig:fig3} is an illustration of how to derive photometric
redshifts from $P_i$.  We display $P_i$ as a function of redshift for
four EDR quasars.  The dashed line is the threshold probability level
of $1/N$.  Each spike in $P_i$ which reaches above this threshold
represents one possibility for $z_{\mathrm{phot}}$.  For each quasar,
the most likely photometric redshifts are listed, along with their
ranges and confidence-levels.  The top two quasars in the figure have
only one likely photometric redshift each, while for the bottom two,
there are several possible $z_{\mathrm{phot}}$.

\subsection{Summary of Changes Made to the RW01 Algorithm}
\label{sec:improvementsRW01}

Between the publication of RW01 and now, we have made the following
refinements to our original photo-$z$ algorithm:
\begin{itemize}
\item The empirical CZR is parameterized in a more refined way.  It is
now parameterized as a series of four-dimensional multivariate
Gaussians with ellipsoidal cross-sections, one for each redshift bin.
(Modulo a $\sim10$\% tail of red quasars, \markcite{R2003}{Richards}
{et~al.}  (2003) found that the color distribution of SDSS quasars is
relatively Gaussian.)  The variances of the Gaussians (which define
the ``width'' of the CZR) are allowed to change as a function of
redshift.  Also, the covariances of the Gaussians are taken into
account, i.e.: the principal axes of the Gaussians' ellipsoidal
cross-sections are not required to be aligned with the four axes of
color-space.
\item The uncertainties in the observed colors of an individual quasar
are handled more rigorously.  The uncertainty in the $u-g$ color of a
quasar is correlated with the uncertainty in its $g-r$ color (since
both involve the quasar's $g$-magnitude); in other words, the
covariance between the quasar's $u-g$ and $g-r$ colors is not zero.
When finding the photometric redshift of a particular quasar, we
therefore consider the full covariance matrix (covariances as well as
variances) for its observed colors.  As a corollary to these first two
refinements, the expression for the $\chi^2$ statistic is more
complicated, as shown in Section~\ref{sec:algstep3}.
\item The photometric redshift of a quasar is now determined by
maximizing its probability $P_i$, rather than minimizing its $\chi^2$
as a function of redshift.  Now that we allow the width of the CZR to
vary with redshift, it is no longer true that $P_i$ is solely related
to $\chi^2_i$; it also depends on the covariance matrix of the CZR.
For this reason, we can no longer let $\chi^2$ minimization stand in
for probability maximization, but rather we must let the latter
determine the photometric redshift.
\item We can optionally take into account the fact that certain quasar
redshifts are more likely than others, if that knowledge is available
a priori.  For instance, suppose we have a large set of
quasars for which we wish to find photometric redshifts, and for some
small subset of these quasars we have spectra and therefore know their
spectroscopic redshifts.  As long as this subset was chosen in an
unbiased way, it is reasonable to assume that the redshift
distribution of all the quasars is identical to the redshift
distribution of the subset.  Thus, we can weight the photometric
redshift probabilities $P_i$ by the redshift distribution of the
subset, so that certain redshifts will be favored more than others.
\item If a quasar candidate is detected as an extended source by the SDSS
photometric pipeline \markcite{lgi+01}({Lupton} {et~al.} 2001), its redshift is probably
$\lesssim 1$.  In the current photometric
redshift technique, we only consider redshifts less than 1 for the
$z_{\mathrm{phot}}$ of an extended object.
\end{itemize}

\section{Tests of the Algorithm Using Confirmed Quasars}\label{chap:EDRQtests}

\subsection{Efficiency of the Algorithm on Known Quasars}
\label{sec:EDRvsEDR}

To compare the efficiency of the new photometric redshift algorithm
with that of the one from RW01, the algorithm was tested on the Early
Data Release Quasar Catalog \markcite{Schn2002}({Schneider} {et~al.} 2002).  This catalog
consists of 3814 quasars with both SDSS photometry and spectra.
Unlike the set of 2625 quasars and Seyferts that we used to test the
earlier version of the algorithm in RW01, all of the EDR quasars have
at least one emission line with FWHM $> 1000$ km s$^{-1}$, a
luminosity of $M_{i}<-23$, and very reliable redshifts.  (It should be
noted, however, that neither data set is statistically homogeneous,
since the quasar-selection algorithm was still undergoing changes and
improvements while these data-sets were being made; see
\markcite{slb+02}{Stoughton} {et~al.} 2002.)  Although there exist newer data than the EDR data,
the number of objects with redshifts and the quality of their
photometry are more than sufficient for our tests.  Furthermore,
similar results are obtained for a much larger sample of DR1 objects
analyzed by \markcite{R2004}{Richards et al.} (2004).

\subsubsection{Tests where $\mathcal{S}$ and $\mathcal{P}$ are identical}
\label{sec:S=P}

After constructing a CZR from the 3814 EDR quasars, the photo-$z$
algorithm was run on the same set of quasars; these photometric
redshifts were compared to the true, spectroscopic redshifts.  Since,
for these tests, $\mathcal{S}$ and $\mathcal{P}$ are identical, we
used $W_i$ to weight $P_i$ by the redshift distribution of the EDR
quasars.

Figures~\ref{fig:fig4} and~\ref{fig:fig5} show the results of this
exercise.  In Figure~\ref{fig:fig4}, we plot the most probable
photometric redshift $z_{\mathrm{phot}}$ of each quasar against its
spectroscopic redshift $z_{\mathrm{spec}}$.  Most of the photometric
redshifts agree well with the spectroscopic redshifts, with incorrect
photometric redshifts generally coming in the form of degeneracies
between very different redshifts (e.g., between $z=0.6$ and $z=1.6$)
or from ``smearing'' of nearby redshifts (e.g.,
$z_{\mathrm{phot}}\sim1.4$).  Of the 3814 quasars, the number
(percent) with photometric redshifts that were correct to within 0.1,
0.2, and 0.3 were: 2348 (61.6\%), 2963 (77.7\%), and 3159 (82.8\%),
respectively.  This result should be compared to the corresponding
result in RW01 (using the old version of the algorithm), where only
70.0\% of the quasars' photometric redshifts were correct to within
0.2.  In Figure~\ref{fig:fig5}, we plot a histogram of $\Delta z =
z_{\mathrm{spec}}-z_{\mathrm{phot}}$ for these same quasars.

As discussed in Section~\ref{sec:algstep3}, our photometric redshift
algorithm supplies not only a value for each quasar's
$z_{\mathrm{phot}}$, but a range around that $z_{\mathrm{phot}}$, and
an associated confidence-level (the probability that the spectroscopic
redshift is within the range).  For example, the photometric redshift
of a quasar might be given as $2.35^{+0.10}_{-0.15}$, to 85\%
confidence.  Of the 3814 EDR quasars, the spectroscopic redshift lies
within the range for 3176 of them (83.3\%).  Figure~\ref{fig:fig6}
shows the distribution of range sizes.  The median range size is 0.42,
and 90\% of the ranges are smaller than 0.64.  Thus most of our
$z_{\mathrm{phot}}$ error bars are $\pm 0.3$ or smaller.

We test the confidence level supplied with the photometric redshifts
in the following manner.  Consider all of the quasars whose
photometric redshift ranges have associated confidence levels of (say)
50-55\%.  One would expect that roughly 50-55\% of these quasars would
have their spectroscopic redshifts within the given range, if the
confidence level was reliable.  Figure~\ref{fig:fig7} shows that the
confidence levels pass this test very well, especially when there are
many quasars with the same confidence level: roughly $x$\% of the
objects with $x$\% confidence level have a spectroscopic redshift
within the range.

\subsubsection{Tests where $\mathcal{S}$ and $\mathcal{P}$ are different}
\label{sec:S!=P}

In any practical application of the photometric redshift algorithm,
$\mathcal{S}$ and $\mathcal{P}$ will not be identical sets.  We
therefore performed two more tests using the EDR quasars, to measure
the efficiency of the method in this more realistic case.

In the first test, we randomly chose 954 (25\%) of the EDR quasars
from which to construct the CZR.  This CZR was used to predict
photometric redshifts for the remaining 2860 EDR objects.  Since in
this case $\mathcal{S}$ and $\mathcal{P}$ are both drawn from the EDR
quasars, the redshift distributions of the two sets will be the same.
So for this test, we weighted the photometric redshifts of the quasars
in $\mathcal{P}$ by the redshift distribution of the quasars in
$\mathcal{S}$.  Of the 2860 quasars, the number (percent) of them with
photometric redshifts correct to within 0.1, 0.2, and 0.3 are: 1598
(55.9\%), 2136 (74.7\%), and 2312 (80.8\%), respectively.  The
efficiency of the method is nearly as high as when the quasars in the
CZR and the quasars being tested were identical.

In the second test, we made the CZR from the 2829 EDR quasars with
$i<19$, and used it and our $z_{\mathrm{phot}}$ technique to obtain
photometric redshifts for the 985 $i>19$ objects.  For this test, the
redshift distributions of $\mathcal{S}$ and $\mathcal{P}$ are not
necessarily the same.  Therefore, we did not weight the photometric
redshifts by the redshift distribution of the CZR.  Of the 985 EDR
quasars with $i>19$, the number (percent) of them with photometric
redshifts correct to within 0.1, 0.2, and 0.3 are: 499 (50.7\%), 681
(69.1\%), and 733 (74.4\%), respectively.  These efficiencies are
somewhat lower than those of the previous tests, because all of the
objects in $\mathcal{P}$ are faint, and thus have larger photometric
uncertainties.  Nevertheless, the results are still quite good
compared with previous work in this field.

\subsubsection{Tests Using Multiple CZRs}

Another test worth mentioning is one where instead of comparing the
colors of input objects to a single CZR, we compared colors to
multiple CZRs defined as a function of their broad spectral slope.
Specifically, we created multiple CZRs for quasars as a function of
their $\Delta(g-i)$ colors.  We can then determine photometric
redshifts with respect to each of these CZRs.  This approach is
motivated by the finding of \markcite{R2003}{Richards} {et~al.} (2003) that quasars with different
optical colors have somewhat different emission line properties.
Unfortunately, we found that using multiple CZRs did not help and we
will hereafter only consider photometric redshifts derived from
comparisons to the single median CZR.

\subsection{Problematic Quasars}
\label{sec:challenges}

\subsubsection{Extended Quasars}\label{sec:extend}

In Figure~\ref{fig:fig4}, the 122 black squares are quasars that were
classified as ``extended'' (i.e., not a point source) by the SDSS
photometric pipeline \markcite{lgi+01}({Lupton} {et~al.} 2001) based on a comparison of PSF and
model magnitudes.  Most of these objects have $z_{\mathrm{spec}}<1$ as
expected, justifying our weighting of extended sources to low
redshift.  Extended objects with spectroscopic redshifts larger than 1
are most likely either lensed quasars, star-quasar superpositions, or
galaxy-quasar superpositions.  Any of these, if unresolved, could be
mistaken for a single extended source.

Our use of PSF magnitudes will partially mitigate the effects of the
host galaxy, but the PSF quasar photometry is still influenced by host
galaxy light for extended sources.  Host galaxy light will alter the
observed colors of the quasars, which would cause a problem with
obtaining accurate photometric redshifts.  Indeed, when the earlier
version of our $z_{\mathrm{phot}}$ algorithm (presented in RW01) was
applied to the EDR quasars, 46 out of the 122 extended quasars
(37.7\%) had $|\Delta z|>0.3$; most of these were assigned photometric
redshifts greater than one.

By restricting the redshift range of extended sources, there are now
only 19 extended quasars with incorrect photometric redshifts.  Since
16 have $z_{\mathrm{spec}}>1$, we are really missing only three
quasars.  (Of course, the 27 extended quasars recovered in this way
are negligible compared to the 3814 quasars in the entire sample, so
our overall efficiency only increased by a fraction of a percent.)
Thus, it appears that we do not have to worry about getting incorrect
photometric redshifts for extended quasars.

\subsubsection{Reddened Quasars}\label{sec:reddened}

Of the 3814 quasars in the EDR, 313 are reddened according to our
definition; these are plotted with black pluses in
Figure~\ref{fig:fig4}.  Since these quasars have unusual colors for
their redshifts, it is clear that their photometric redshifts will
probably be incorrect.  An examination of Figure~\ref{fig:fig4} shows
that this is indeed the situation; of the 313 reddened quasars, 192 of
them (61.3\%) have photometric redshifts with $|\Delta z|>0.3$.

Since the majority of reddened quasars will have incorrect
$z_{\mathrm{phot}}$, we could increase the efficiency of our
photometric redshift algorithm if they could be identified from
photometric information alone, and excluded from the sample.
Figure~\ref{fig:fig8} shows the positions in color-space of the
reddened quasars (black points) and the non-reddened quasars (gray
points).  For the most part, the non-reddened quasars are found in a
four-dimensional ellipsoidal Gaussian, and the reddened quasars are in
a second ellipsoidal Gaussian, substantially offset from the
non-reddened one.  (The quasars not found in the Gaussians are $z
\gtrsim 2.5$ quasars, which have the Lyman $\alpha$ forest redshifted
into the SDSS filters.)

Here we have a similar problem to that of finding photometric
redshifts.  In both cases, quasars of different kinds are found in
roughly ellipsoidal Gaussians, separated from each other in
color-space.  If we quantify the positions, sizes, and shapes of the
different Gaussians, we can calculate the probability that any
particular object (a point in color-space) was drawn from one of the
Gaussian distributions.  Hence, we can use the photometric redshift
technique itself (slightly modified) to obtain an approximate
probability that a given quasar is reddened.

We used the CZR branch of the $z_{\mathrm{phot}}$ algorithm (see
Section~\ref{sec:algstep2}) to measure the ellipsoidal Gaussians for
the reddened and non-reddened quasars.  To remove the $z \gtrsim 2.5$
quasars, we ran ten iterations of the outlier-removing procedure (see
\S~\ref{sec:algstep2}), instead of two.  Projections of the derived
Gaussians are plotted in Figure~\ref{fig:fig8} with a dashed line (for
the non-reddened quasars) and a solid line (for the reddened quasars).

We then used the last part of the $z_{\mathrm{phot}}$ algorithm (see
Section~\ref{sec:algstep3}) to calculate the probability, $P_{nr}$,
that an EDR quasar belongs to the non-reddened distribution, and the
probability, $P_r$, that it belongs to the reddened distribution
(where $P_{nr}+P_r=1$).  This was done for all 3814 EDR quasars.
Since there are 11.2 times more non-reddened quasars than there are
reddened quasars, we weighted $P_{nr}$ and $P_r$ accordingly.

We then can use $P_r$ to select out mostly non-reddened quasars.  If
we consider only the EDR quasars with $P_r<0.99$, we will have 3384
quasars, and only 106 of them (or 3.1\%) will be reddened (a reduction
in total sample size of 13\% but a drop in the ``contaminated
fraction'' from 8.2\% to 3.1\%).  Subsets with a higher fraction of
non-reddened quasars can be made by lowering the upper bound on $P_r$,
although at the cost of making the subsets somewhat smaller; for
example, there are only 2617 quasars with $P_r<0.01$, but only one of
them is reddened.

Although we have not made such $P_r$ cuts in this paper, we see that
$P_r$ can be used to largely eliminate reddened quasars using only
photometric information.  In future work, we will use $P_r$ to improve
the efficiency of our photometric technique still further.

\subsubsection{Color-redshift Degeneracies}\label{sec:degeneracies}

Of the 655 EDR quasars with $|\Delta z| \ge 0.3$, 192 are reddened,
and 19 are extended.  That still leaves 444 ``ordinary'' quasars with
incorrect $z_{\mathrm{phot}}$ --- and many of these are
\textit{really} incorrect, with $|\Delta z|=1.0$ or even $2.0$!

An examination of Figure~\ref{fig:fig4} shows that these quasars are
placed roughly symmetrically about the
$z_{\mathrm{phot}}=z_{\mathrm{spec}}$ diagonal.  For instance, some
quasars with $z_{\mathrm{spec}}=0.8$ are assigned a
$z_{\mathrm{phot}}$ of 2.1, whereas some quasars with
$z_{\mathrm{spec}}=2.1$ are given a $z_{\mathrm{phot}}$ of 0.8.  This
situation arises because there is considerable degeneracy in the
color-redshift relation for $z\lesssim2.5$ --- i.e.: quasars at fairly
different redshifts (like 0.8 and 2.1) can still have very similar
colors.  Color-redshift degeneracies occur when the CZR intersects
itself (or does so very nearly) in color-space --- see the inset of
Figure~\ref{fig:fig2} for an illustration of this.

\begin{figure}[p]
\plotone{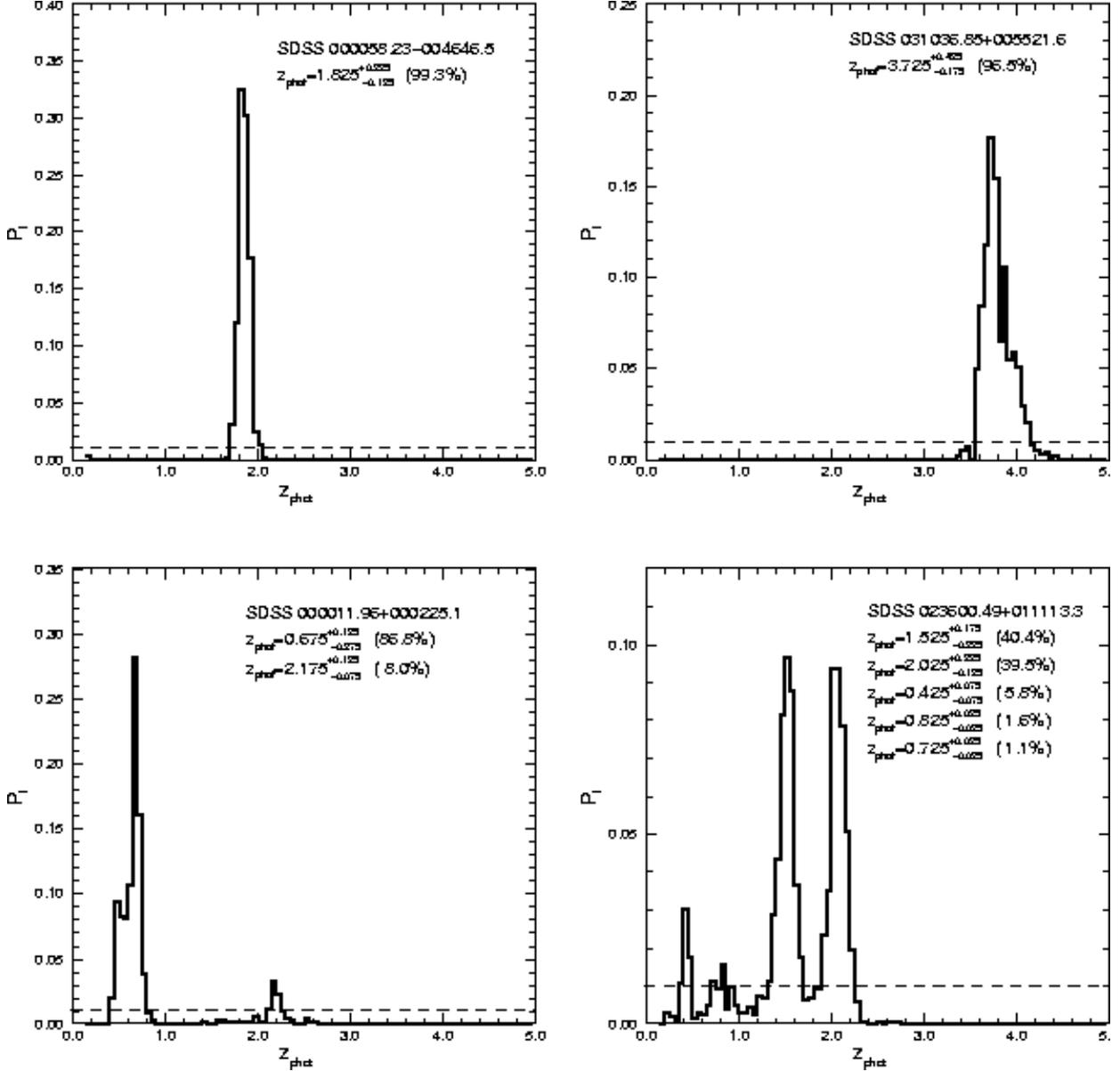}
\caption{The probability distribution ($P_i$) for the photometric
redshifts for four EDR quasars.  The top two panels show objects with
only one likely photo-$z$; the bottom two panels show examples of
objects with more than one likely photo-$z$.  For each quasar, the
most likely photometric redshift(s) are given along with the range and
confidence-levels.  Their true redshifts are (clockwise from
top-left): 1.896, 3.780, 1.81, and 0.479.}
\label{fig:fig3}
\end{figure}

\begin{figure}[p]
\plotone{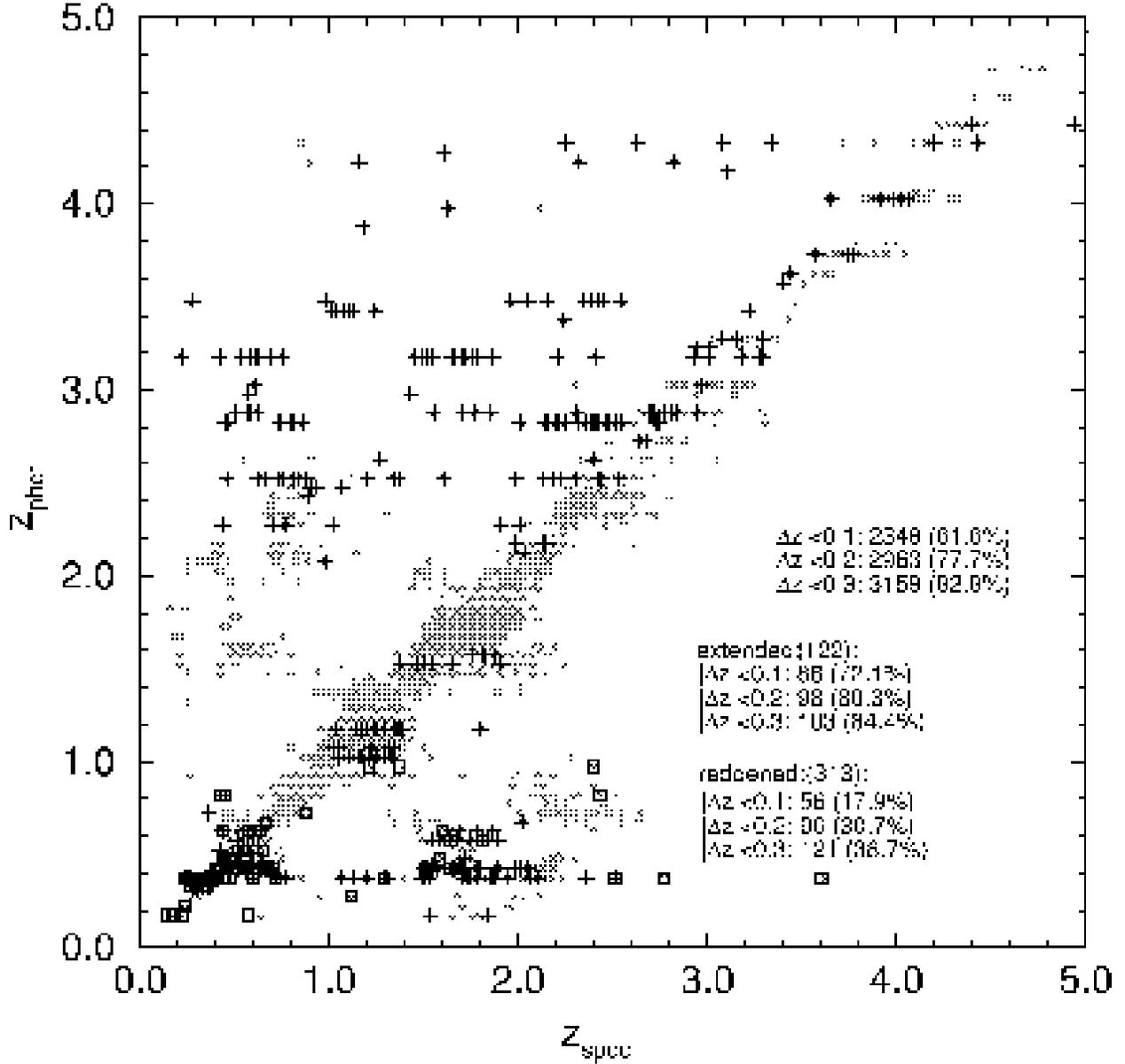}
\caption{Photometric redshift vs.\ spectroscopic redshift 
for the 3814 EDR quasars.  \textit{Gray circles}: non-reddened quasars, 
{\em black pluses}: reddened quasars, {\em black squares}:
extended quasars.}
\label{fig:fig4}
\end{figure}

\begin{figure}[p]
\plotone{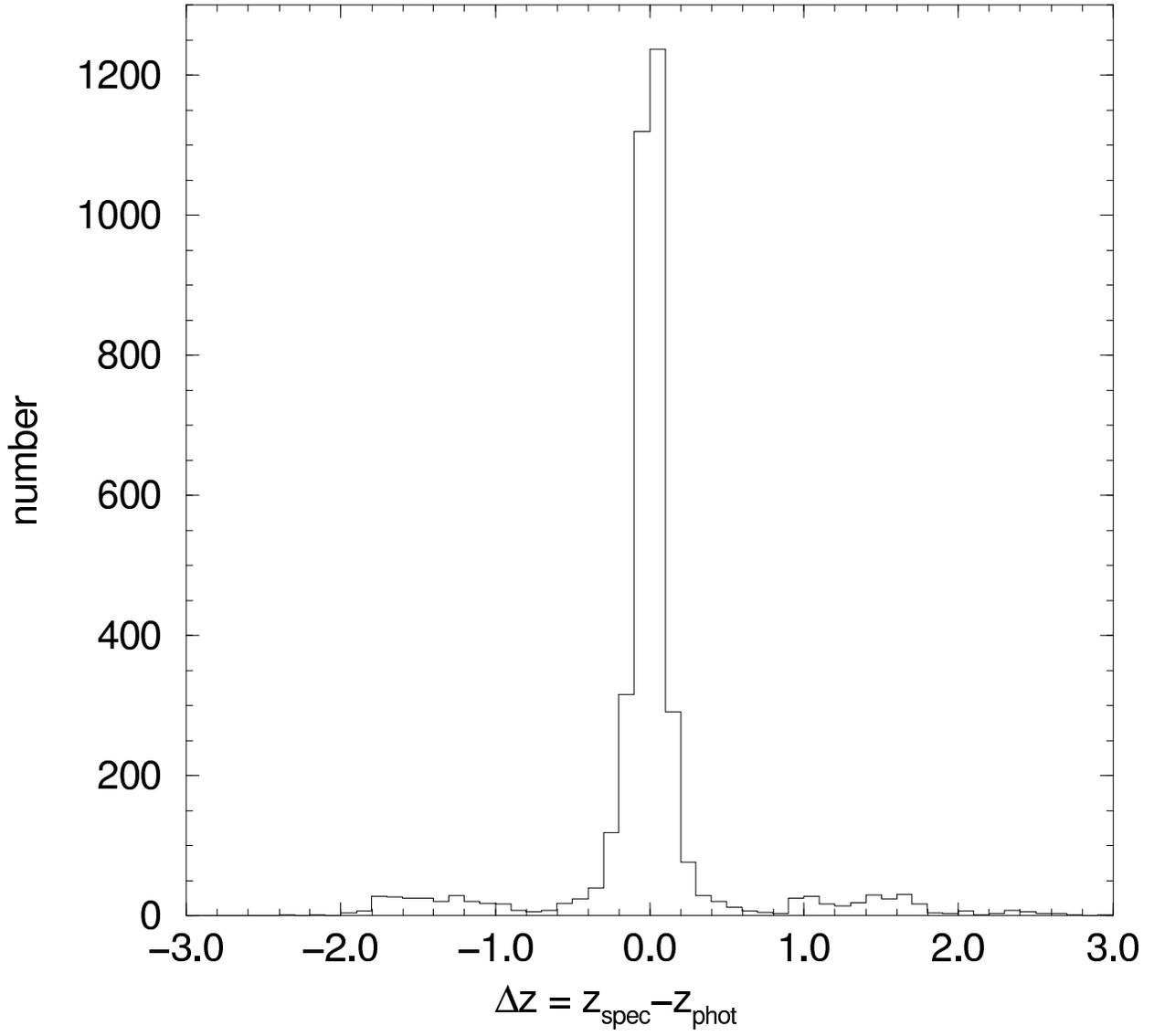}
\caption{Distribution of $\Delta z = z_{\mathrm{spec}}-z_{\mathrm{phot}}$ for EDR quasars.}
\label{fig:fig5}
\end{figure}

\begin{figure}[p]
\plotone{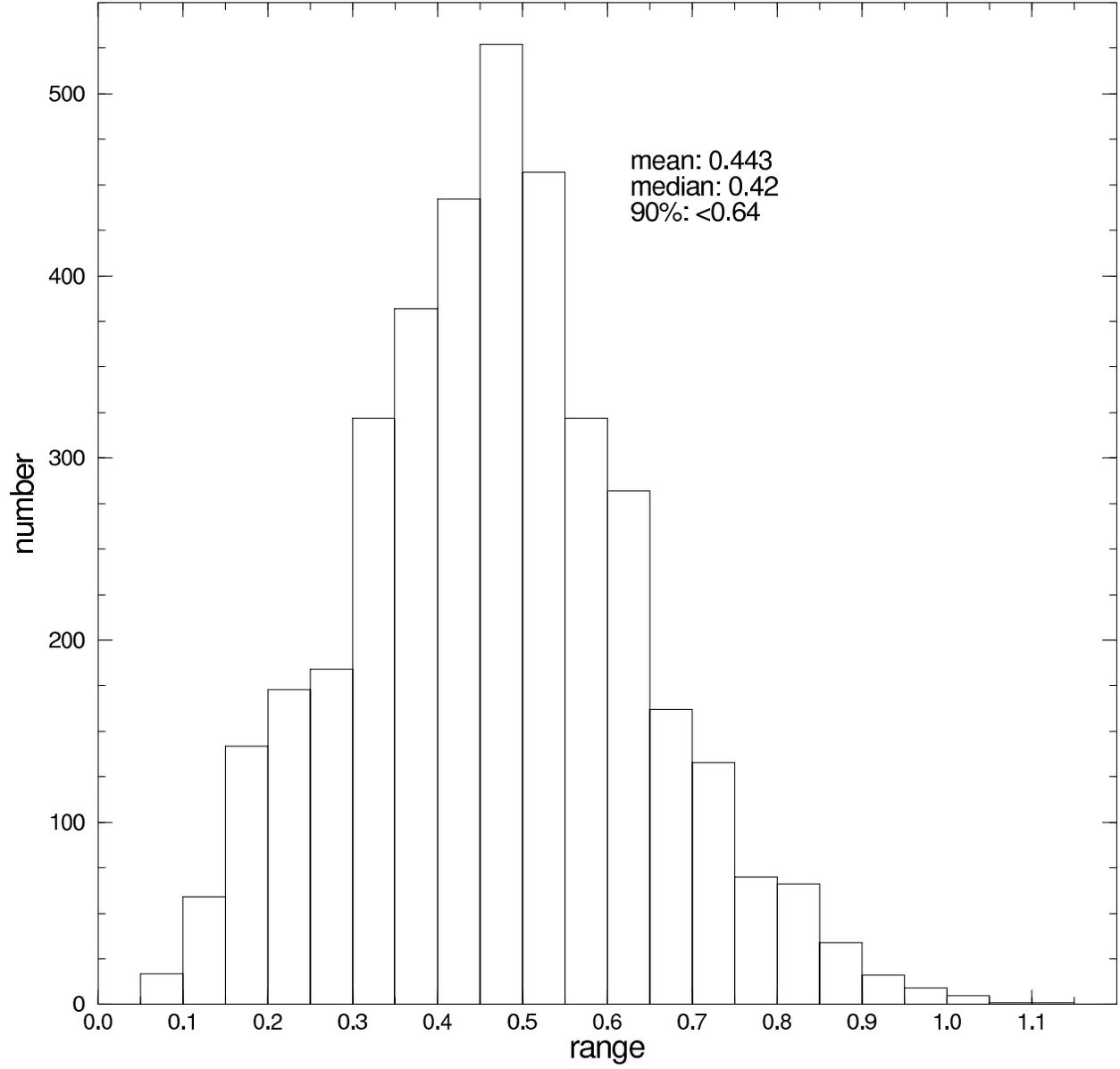}
\caption{Distribution of photometric redshift range sizes for the EDR
quasars.}
\label{fig:fig6}
\end{figure}

\begin{figure}[p]
\plotone{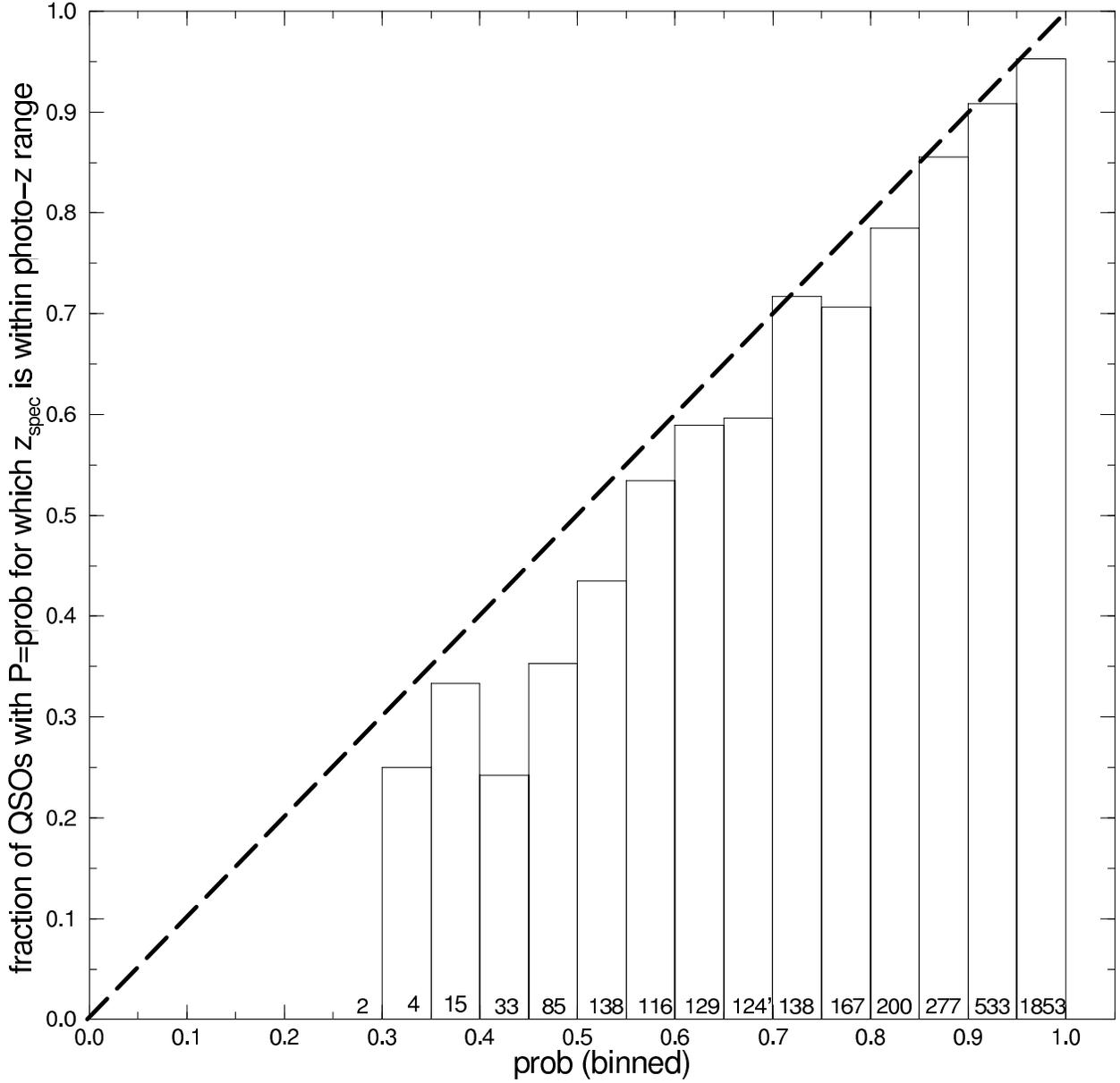}
\caption{Fraction of EDR quasars with confidence-level $P$ equal to
\textit{prob} for which $z_{\mathrm{spec}}$ is within the photo-$z$
range, where \textit{prob} goes from 0 to 1 in bins of 0.05.
\textit{Numbers} denote total number of EDR quasars with
confidence-level $P$.}
\label{fig:fig7}
\end{figure}

\begin{figure}[p]
\plotone{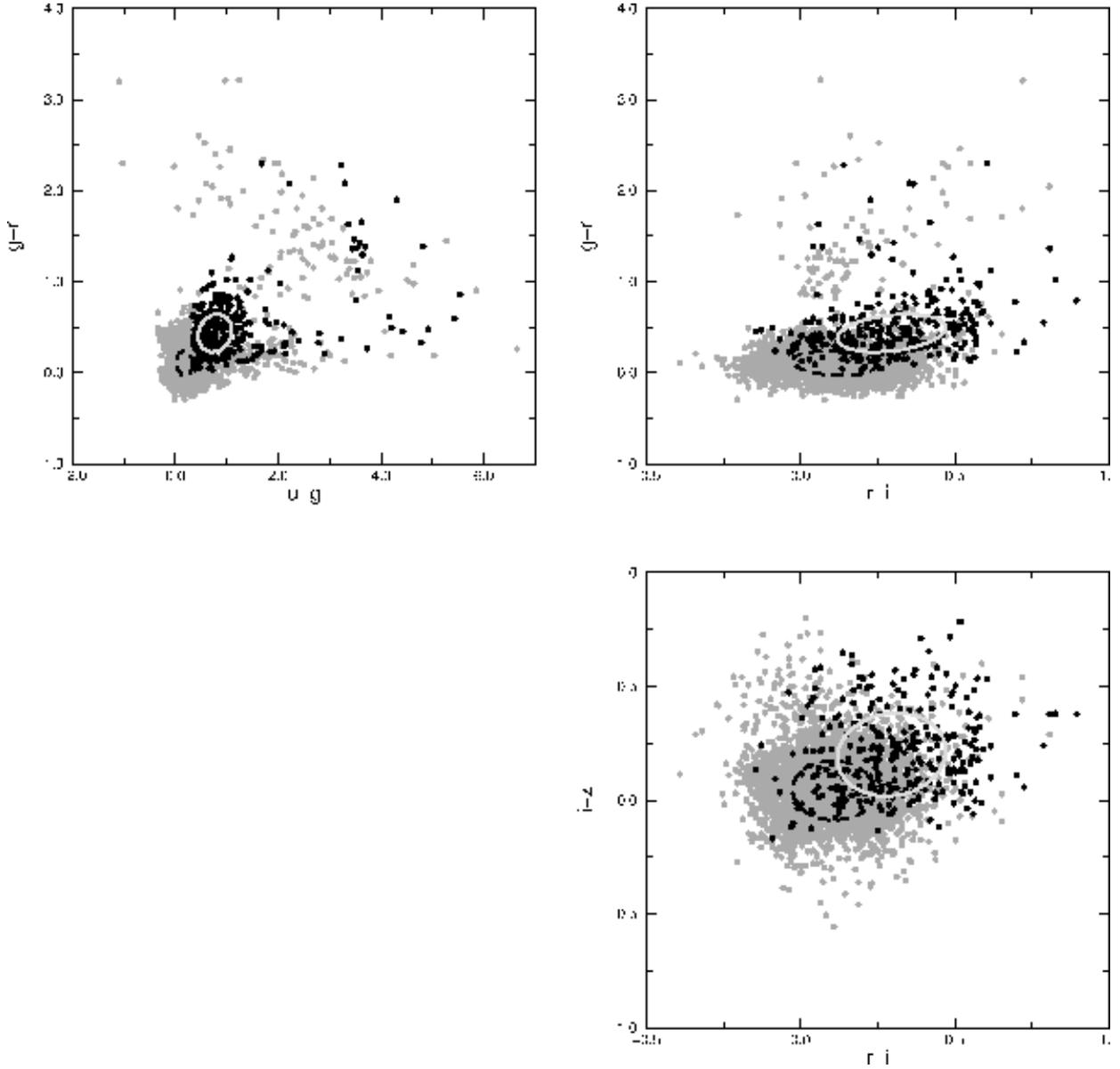}
\caption{Distributions in color-space of non-reddened EDR quasars 
\textit{(gray points)} and reddened EDR quasars \textit{(black points)}.  
\textit{Dashed black ellipses}: projections of 1-$\sigma$ ellipsoid for 
non-reddened quasar distribution; \textit{solid gray ellipses}: projections of 
1-$\sigma$ ellipsoid for reddened quasar distribution.}
\label{fig:fig8}
\end{figure}

When a quasar's colors are consistent with two (or more) redshifts,
the function $P_i$ (the probability that its redshift is in redshift
bin $i$) has two or more peaks of roughly equal size, at each of the
possible redshifts.  An example is the bottom-right plot in
Figure~\ref{fig:fig3}, which shows a degeneracy between $z=1.5$ and
$z=2.0$ (easily seen in Figure~\ref{fig:fig4}).  If the algorithm is
looking for the most probable $z_{\mathrm{phot}}$, it may or may not
pick the correct one.  Fortunately, our algorithm can be directed to
search out the $n$ most likely photometric redshifts, where $n$ is any
desired integer.  Thus, if we find that two or more choices for
$z_{\mathrm{phot}}$ have similar confidence-levels, we at least know
that one of them is probably correct (although we do not know which
one!).

If $P_i$ is being weighted by the redshift distribution of the CZR,
that may help the algorithm select the correct $z_{\mathrm{phot}}$.
For instance, in the EDR, there are twice as many quasars with
$z_{\mathrm{spec}}=1.5$ as there are with $z_{\mathrm{spec}}=2.0$.  So
even though quasars at either redshift have very similar colors, a
quasar with these colors is twice as likely to have a redshift of 1.5,
and the weighting of $P_i$ will reflect this.  Of course, this will
only work if the redshift distribution of $\mathcal{P}$ is similar to
that of $\mathcal{S}$.  It should also be noted that this weighting
technique will be of most use if one is searching for the $n$ most
likely photometric redshifts, and listing them by probability (as is
done in Fig.~\ref{fig:fig3}).  If one is only searching for the most
likely photometric redshift, then the weighting will cause the
algorithm to always select the more common redshift when faced with a
true degeneracy (e.g., $z=1.5$ over $z=2.0$).

\begin{figure}[p]
\plotone{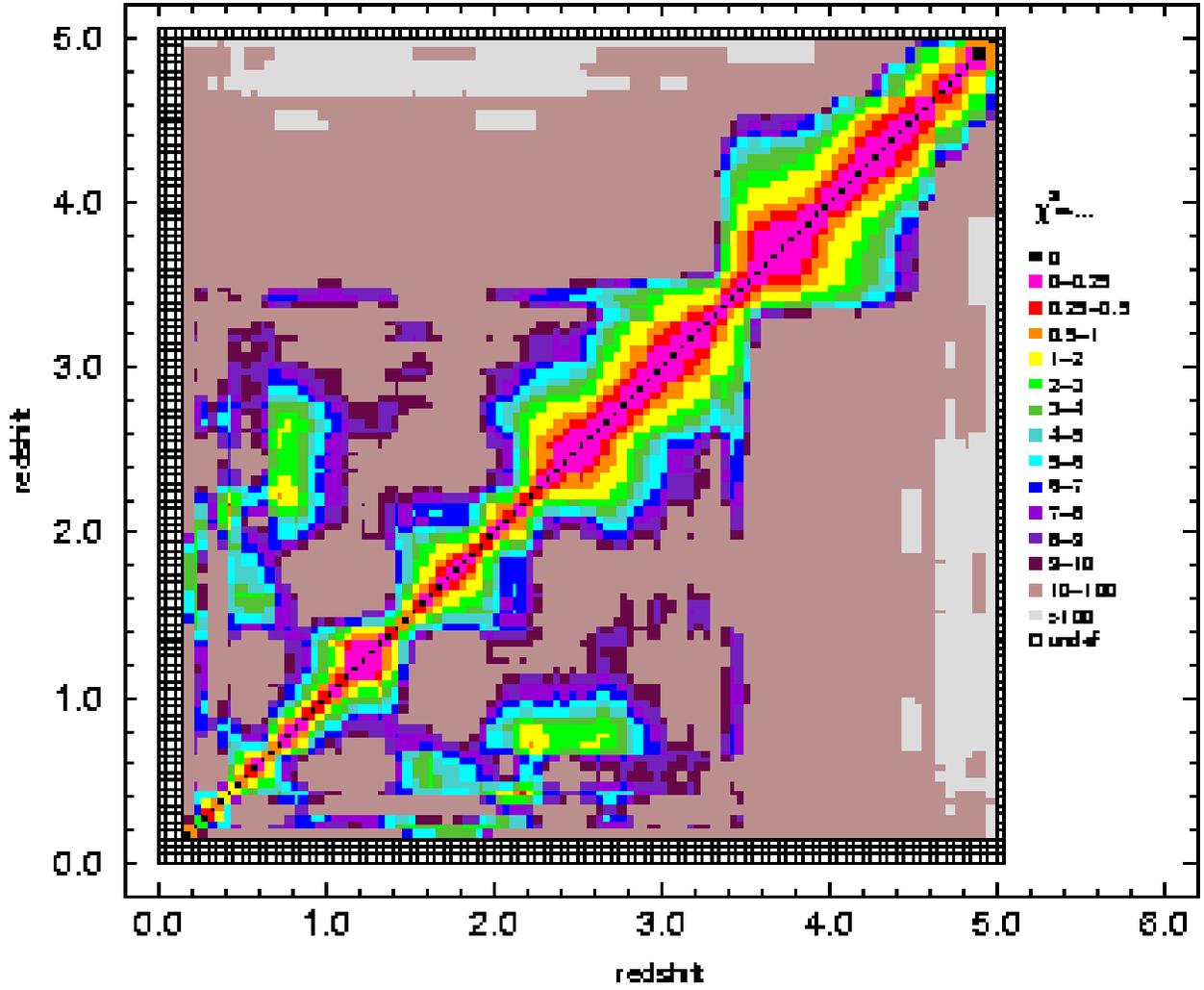}
\caption{Degeneracy map for EDR CZR.  Color of pixel at position 
$(x,y)$ represents $\chi^2$ ``distance'' in color-space between 
redshifts $x$ and $y$.}
\label{fig:fig9}
\end{figure}

Since degenerate redshifts are found where the CZR intersects itself
in color-space, we can create a map of the degenerate redshifts by
measuring the ``distance'' in color-space between any two points on
the CZR --- the closer they are, the more degenerate the corresponding
redshifts.  Figure~\ref{fig:fig9} is such a map.

In Figure~\ref{fig:fig9}, the color of the pixel at position $(x,y)$
represents the $\chi^2$ ``distance'' between redshift $x$ and redshift
$y$ in the CZR.  The smaller the value of $\chi^2$, the more similar
the colors of quasars at these redshifts.  Notice that the structure
of Figures~\ref{fig:fig4} and \ref{fig:fig9} are similar (modulo the
fact that Figure~\ref{fig:fig9} is symmetric by definition, whereas
Figure~\ref{fig:fig4} need not be).  Thus, consulting this map is a
way to determine whether or not a quasar with a certain
$z_{\mathrm{phot}}$ has degenerate colors.

\section{Tests of the Algorithm Using Quasar Candidates}\label{chap:FEDRtests}

In this section, we present a second round of tests, in which we apply
our $z_{\mathrm{phot}}$ algorithm to quasar candidates selected by the
final version of the SDSS quasar-selection algorithm.  We call this
data-set the Final Early Data Release (FEDR), since its objects are
found in the same region of sky as the EDR.  Many of these objects
have no spectra, and a significant fraction of those that do are not
quasars.  We also introduce the concepts of the ``Window'' (a
$z_{\mathrm{phot}}$ cut) and the ``Box'' (a series of color cuts)
through which --- by solely photometric means --- we can select quasar
candidates that are mostly quasars, the majority of which have
accurate photometric redshifts.


Although the EDR quasar tests presented in \S~\ref{chap:EDRQtests} are
encouraging, and show that the photometric redshift algorithm is
working well, they are not realistic exercises in that we already know
that all 3814 objects are quasars.  In a real application, the objects
chosen to be quasar candidates (based on their photometry only) will
include many non-quasars.  The tests described in this section are
designed to see what happens when our $z_{\mathrm{phot}}$ algorithm
operates on more realistic data.  For these tests, our CZR was, as
before, constructed from the EDR Quasar Catalog.  However, we then
used this CZR to find photometric redshifts for the FEDR, a set of
8740 objects which we describe below.  (Since in this case
$\mathcal{S}$ and $\mathcal{P}$ probably do not have the same redshift
distributions, we did not weight the photometric redshifts.)

\subsection{FEDR, and How it Differs from the EDR Quasars}\label{sec:8740info}

The FEDR is a set of quasar candidates from the same region of sky as
the EDR quasars.  The quasar-selection algorithm used to select these
objects was the final, current version (discussed in
\markcite{rfn+02}{Richards} {et~al.} 2002); as such these objects are statistically
homogeneous, in contrast to the EDR quasars from \markcite{Schn2002}{Schneider} {et~al.} (2002).
Since the FEDR objects were selected solely by photometric means, not
all have spectra.

Of the 8740 objects, 3985 are of unknown type.  (Most of these have no
spectra; 22 have unidentifiable spectra.)  The other 4755 were
identified by the SDSS selection algorithm as: 52 ``late-type'' stars
(type M, L, or T), 527 ``regular'' stars, 828 normal galaxies, and
3348 quasars and AGNs.

In presenting the results of the FEDR tests, we first consider how
well the photometric redshift algorithm performed for 1) the 3348 FEDR
objects known to be quasars, 2) the 1407 FEDR objects known to be
non-quasars, and 3) the 3985 FEDR objects of unknown type.

\subsection{Objects Confirmed to be Quasars}\label{sec:confirmedquasars}

Of the 3348 objects spectroscopically confirmed to be quasars, the
number (percent) with photometric redshifts that were correct to
within 0.1, 0.2, and 0.3 were: 2156 (64.4\%), 2692 (80.4\%), and 2860
(85.4\%), respectively.  There were 2783 quasars (83.1\%) whose
spectroscopic redshifts lay within the photometric redshift ranges.
The distribution of range sizes has a median of 0.39, and 90\% of the
ranges are smaller than 0.64.  The confidence levels quoted for the
photometric redshifts' error bars appear to be reasonable.  These
results are quite similar to those found for the EDR quasars, not
surprisingly given the degree of overlap in the two sets.

\subsection{Objects that are Confirmed Non-Quasars}
\label{sec:confirmed.nonquasars}

\begin{figure}[p]
\plotone{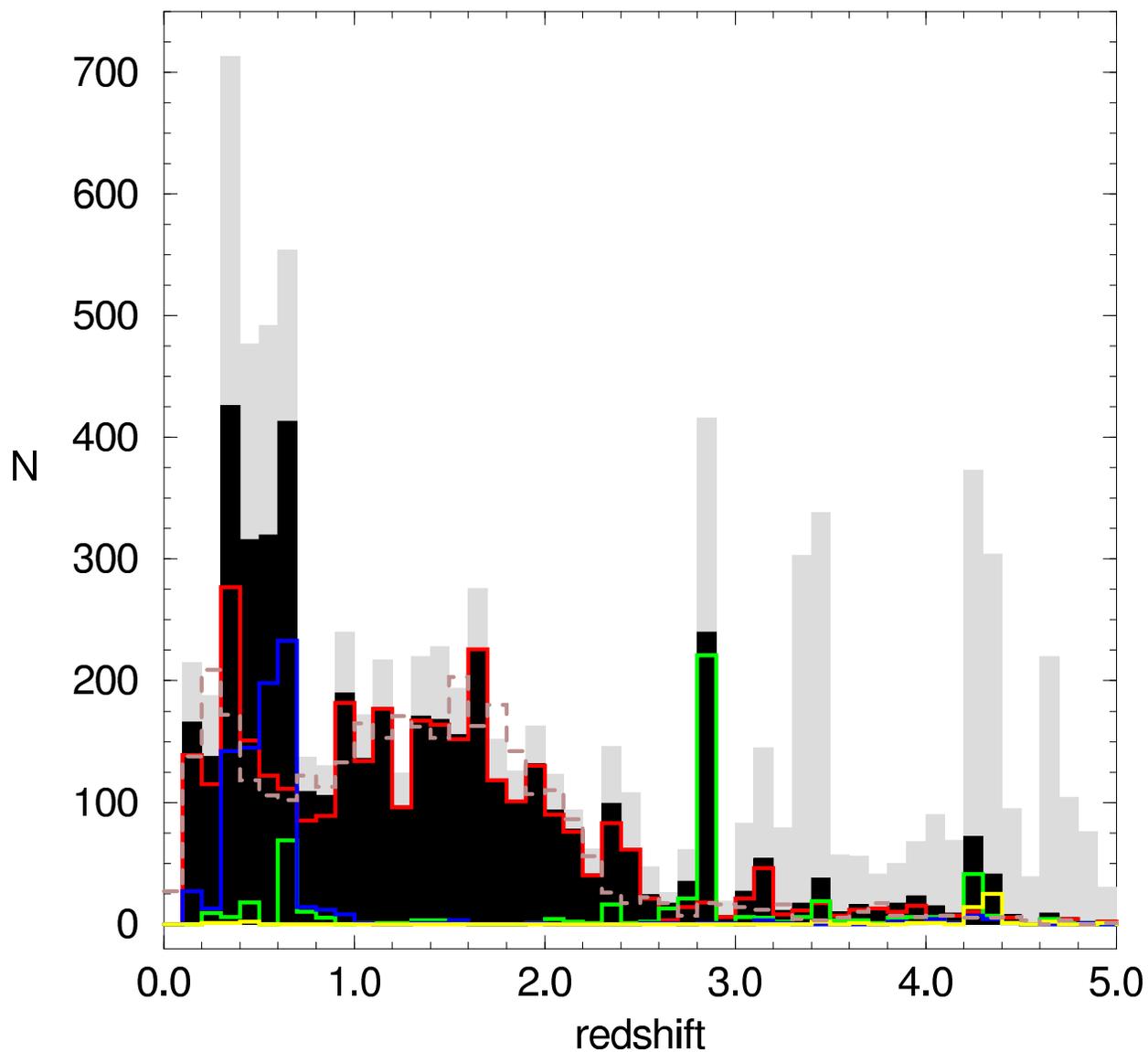}
\caption{Distribution of photometric redshifts for the following 
types of objects in FEDR --- all objects \textit{(gray)}, objects with 
spectra \textit{(black)}, quasars \textit{(red)}, stars 
\textit{(green)}, late-type stars \textit{(yellow)}, galaxies 
\textit{(blue)}.  \textit{Brown, dashed histogram} is distribution 
of spectroscopic redshifts for FEDR quasars.}
\label{fig:fig10}
\end{figure}

Figure~\ref{fig:fig10} is a histogram of the photometric redshifts for
all 8740 objects.  Grey bars show the $z_{\mathrm{phot}}$ distribution
for all of the objects, including those that have no spectra, while
black bars show the $z_{\mathrm{phot}}$ distribution for only the 4755
objects with spectra.  The red, green, yellow, and blue histograms show
the $z_{\mathrm{phot}}$ distributions for confirmed quasars, stars,
late-type stars, and galaxies, respectively.  The brown, dashed
histogram shows the $z_{\mathrm{spec}}$ distribution of the 3348
confirmed quasars, for comparison.

Most of the photometric redshifts assigned to confirmed non-quasars
take on a few specific values.  Nearly all galaxies are assigned
photometric redshifts between 0.3 and 0.7.  Many stars receive a
photometric redshift between 2.8 and 2.9; most others are found
between 0.6 and 0.7, or between 4.2 and 4.3.  Almost all late-type
stars have a photometric redshift between 4.2 and 4.4.  The
color-space distributions of stars and galaxies intersect (or at least
approach) the quasar CZR at these specific redshift values.  Also,
there are almost no non-quasars assigned photometric redshifts between
0.8 and 2.2.

These results are very encouraging.  Suppose one uses the SDSS
quasar-selection algorithm to find quasar candidates.  The majority of
those that receive photometric redshifts between 2.8 and 2.9 are
actually stars (as expected given that the quasar color track crosses
the stellar locus near these redshifts and that the density of stars
far exceeds the density of quasars).  Most other non-quasar
contaminants will have photometric redshifts in the ranges 0.3--0.7,
or 4.2--4.4.  Any objects with these photometric redshifts should be
flagged as suspected non-quasars, while the others (especially those
with $0.8<z_{\mathrm{phot}}<2.2$) are likely to be quasars.

\begin{figure}[p]
\plotone{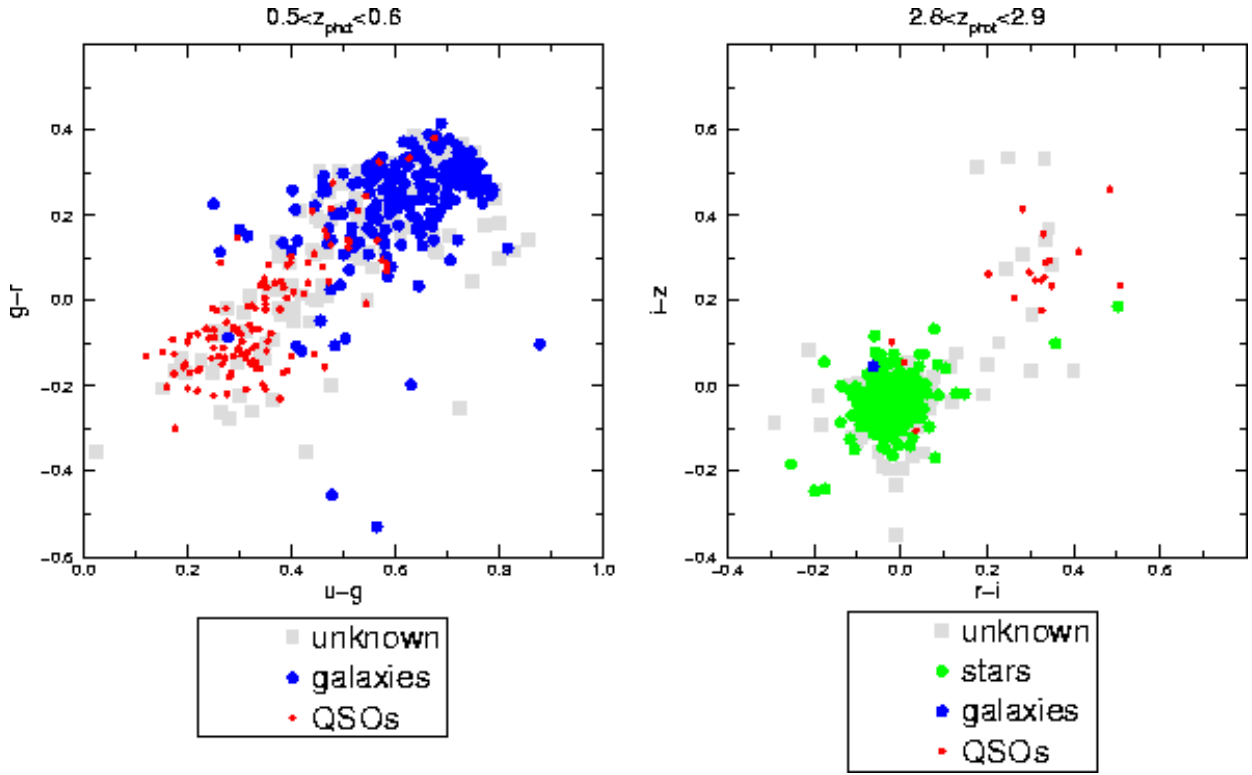}
\caption{Two-dimensional projections of color-space, showing locations
of various classes of objects in the FEDR.  \textit{Left:} FEDR
objects with $0.5<z_{\mathrm{phot}}<0.6$, \textit{Right:} FEDR objects
with $2.8<z_{\mathrm{phot}}<2.9$.  In both plots, \textit{gray
squares} are objects without spectra, \textit{red circles} are
quasars, \textit{green circles} are stars, and \textit{blue circles}
are galaxies.}
\label{fig:fig11}
\end{figure}

We can perhaps do even better by investigating whether non-quasars and
quasars with similar $z_{\mathrm{phot}}$ can be separated by their
colors.  The two plots in Figure~\ref{fig:fig11}, which are
two-dimensional projections of color-space, show the locations of
various objects.  Grey squares are of unknown type.  Red, green, and
blue circles are confirmed quasars, stars, and galaxies, respectively.
The objects in the two plots of Figure~\ref{fig:fig11} have
photometric redshifts in the ranges 0.5-0.6 (\textit{left}), and
2.8-2.9 (\textit{right}).  We see that most quasars and galaxies with
photometric redshifts of 0.5-0.6 can be separated by $u-g$ and $g-r$,
while most quasars and stars with photometric redshifts of 2.8-2.9 can
be differentiated by $r-i$ and $i-z$.  This allows us to retrieve most
quasars with these photometric redshifts.

\subsection{Objects of Unknown Type}\label{sec:unknown}

Referring again to Figure~\ref{fig:fig10}, compare the
$z_{\mathrm{phot}}$ distributions for all objects (\textit{grey bars})
and objects with identifiable spectra (\textit{black bars}).  For
$z_{\mathrm{phot}}<3$, the two distributions are roughly the same
(except for overall normalization).  However, for
$z_{\mathrm{phot}}>3$, and especially for $z_{\mathrm{phot}} \sim 3.5$
and $z_{\mathrm{phot}} \sim 4.5$, there are many more objects without
spectra than we might expect.  This is because of the way the final
quasar-selection algorithm was designed.

Previous versions of the selection algorithm missed many quasars with
$z \sim 3.5$ and $z \sim 4.5$, because these lie in the same region of
color-space as late-type stars.  The final version was therefore
designed to probe deeper into this region \markcite{rfn+02}({Richards} {et~al.} 2002).  Generally
speaking, since the previous versions of the selection algorithm were
used to make the EDR quasar catalog (for which spectra were taken),
most of the objects in the FEDR that do not have spectra were only
selected with the final selection algorithm.  Thus, we find a striking
abundance of objects with no spectra having $z_{\mathrm{phot}}>3$.
Most of these are probably late-type stars that are included as
contaminants in the SDSS's attempt to be as complete as possible to
bright $z>3$ quasars; many of these candidates are from a small number
of data frames that have sub-standard photometry --- causing stars
within the stellar locus to scatter into the quasar distribution.

\subsection{Objects with $0.8<z_{\mathrm{phot}}<2.2$ (``the Window'')}
\label{sec:Window}

We now turn our attention to the objects whose photometric redshifts
lie in the largely uncontaminated window between 0.8 and 2.2.  These
will be called Window objects.

There are 2459 objects in the Window; 1902 of them are confirmed
quasars, while only 22 and 27 are confirmed stars and galaxies,
respectively.  The remaining 508 objects are of unknown type.  If we
assume that the distribution of unknown objects is the same as that of
the known ones, then approximately 97.5\% of the Window objects are
quasars.  If an object selected as a quasar-candidate by the selection
algorithm has a $z_{\mathrm{phot}}$ in the Window, it is almost
certainly a quasar.

The efficiency of the $z_{\mathrm{phot}}$ algorithm is also better in
the Window.  Out of the 1902 confirmed quasars in the Window, the
number (percent) with photometric redshifts correct to within 0.1,
0.2, and 0.3 are: 1314 (69.1\%), 1638 (86.1\%), and 1746 (91.8\%).
Thus, 89.5\% of the objects in the Window (that is, 91.8\% of 97.5\%)
are quasars with photometric redshifts correct to within 0.3.

\subsection{Color-space Cuts that Select Mostly Window Objects (``The Box'')}
\label{sec:Box}

It is also of interest to determine regions of color space that yield
higher than average quasar efficiencies (quasars : quasar candidates).
\markcite{R2004}{Richards et al.} (2004) describe the application of a complex algorithm to the
SDSS-DR1 dataset; however, for many applications users may prefer a
simpler method.  As such, we define a region of color space that was
designed to yield mostly Window objects such that one can dispense
with the quasar-selection algorithm altogether and simply query the
SDSS database for objects with colors in this region of color-space.

\begin{figure}[p]
\plotone{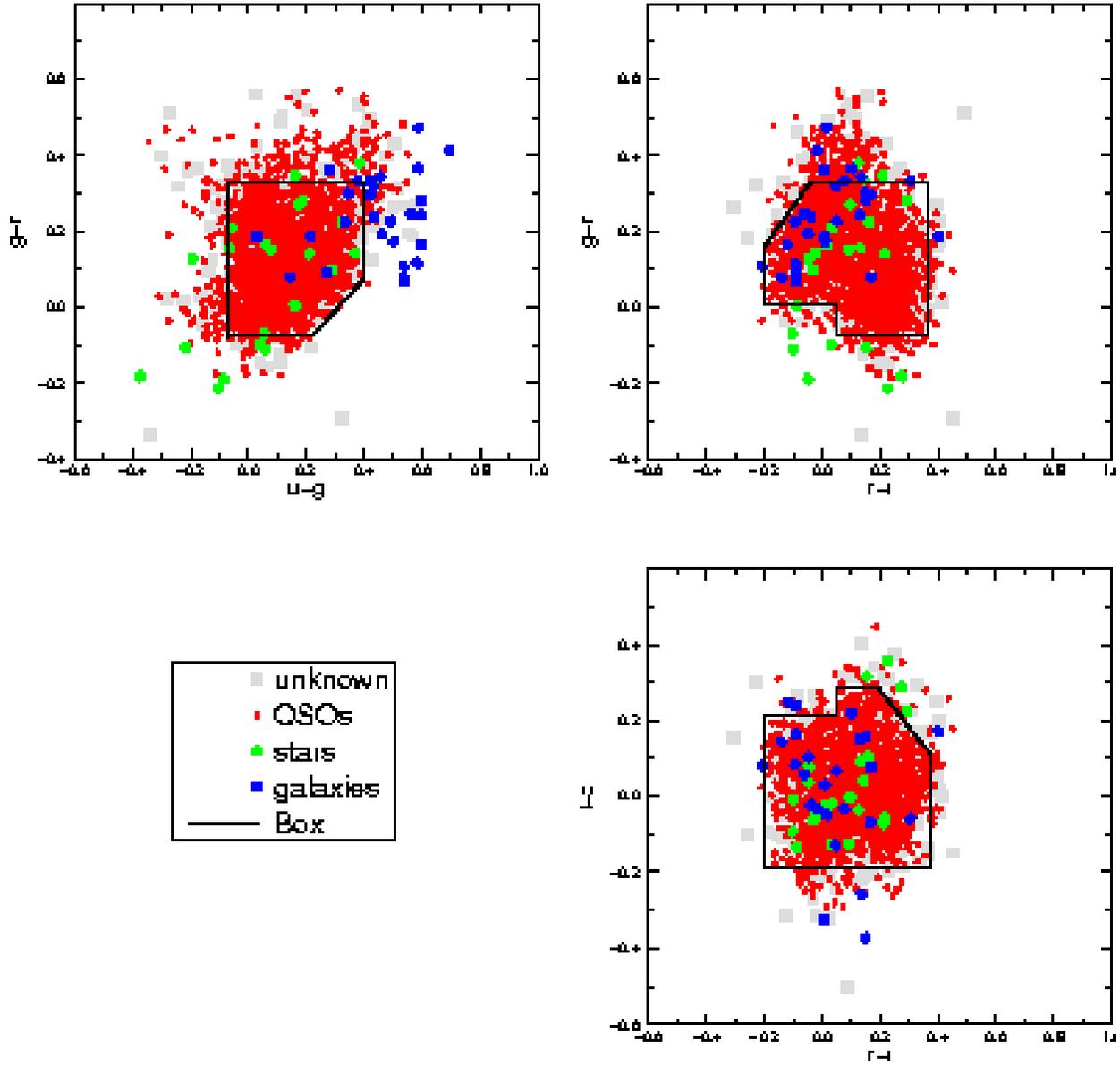}
\caption{Two-dimensional projections of color-space, showing Box 
color-cuts \textit{(black lines)}, and the FEDR objects with 
$0.8<z_{\mathrm{phot}}<2.2$.  \textit{Symbols} are the same as in 
Figure~\ref{fig:fig11}.}
\label{fig:fig12}
\end{figure}

Figure~\ref{fig:fig12} shows a region of color-space which we call
``the Box,'' designed to include primarily Window objects.  The Box is
defined by the intersection of the following cuts in color-space:
\[
\begin{array}{c}
    \begin{array}{rcccl}
        -0.07 & < & u-g & < & 0.40 \\
        \{ 0.01 | -0.075 \} & < & g-r & < & 0.33 \\
        -0.20 & < & r-i & < & 0.37 \\
        -0.19 & < & i-z & < & \{ 0.215 | 0.29 \}
    \end{array} \\
    \begin{array}{rcl}
        g-r & > & 0.8333(u-g) - 0.2583 \\
        g-r & < & 1.0625(r-i) + 0.3725 \\
        i-z & < & -0.9730(r-i) + 0.4749
    \end{array}
\end{array}
\]
where $\{ x | y \}$ is equal to $x$ if $r-i < 0.05$ and $y$ if $r-i
\geq 0.05$.

Of the 8740 objects in our sample, 1752 of them lie in the Box.  1578
of these are objects with $0.8<z_{\mathrm{phot}}<2.2$: 1256 confirmed
quasars, 11 confirmed non-quasars, and 311 unknown objects.  If we
assume that the unknown objects contain the same fraction of quasars
as the known ones, then 89.3\% of the objects in the Box are quasars
with $0.8<z_{\mathrm{phot}}<2.2$.  Thus, the color-cuts we have chosen
do succeed in selecting mostly Window quasars.

Of the 1256 Window quasars found in the Box, most have accurate
photometric redshifts (as expected).  There are 874 with
$z_{\mathrm{phot}}$ accurate to within 0.1, 1086 to within 0.2, and
1159 to within 0.3.  So if we apply the Box color-cuts to SDSS data,
roughly 83\% of them will be quasars with $z_{\mathrm{phot}}$ accurate
to within 0.3.  However, it must be kept in mind that although the Box
is very efficient at picking out mostly Window quasars and little
else, it by no means selects all of the Window quasars.  Within our
sample, there are 1902 confirmed quasars with
$0.8<z_{\mathrm{phot}}<2.2$, whereas the Box contains only 1256, or
66\%.

\section{Summary}\label{chap:summary}

We have presented an updated version of the empirical quasar photo-$z$
algorithm originally presented by RW01.  This improved version of the
code yields redshifts for known quasars that are accurate to $\pm0.3$
in redshift for 83\% of quasars in the EDR quasar catalog.  In
addition, the algorithm returns accurate probabilities that these
redshifts are within a given range --- making it possible to better
identify those 17\% of quasars that have erroneous photo-$z$'s.  We
have further shown that it is possible to use the algorithm to
identify non-quasars among quasar candidates and therefore to
construct the most robust samples (in terms of quasar identification
and photo-$z$ accuracy) among samples of unknown objects.
\markcite{R2004}{Richards et al.} (2004) discuss the application of this photo-$z$ algorithm to a
sample of over 100,000 SDSS-DR1 quasar candidates.  The efficiency of
their selection algorithm combined with the accuracy of our photo-$z$
algorithm suggest that over 80,000 of these objects will be quasars
with accurate photometric redshifts, demonstrating that it will be
possible to construct a sample of hundreds of thousands of quasars by
the end of the SDSS project.  The C code for the algorithm described
herein can be obtained by contacting either of the first two authors.

\acknowledgements

Funding for the creation and distribution of the SDSS Archive has been
provided by the Alfred P.  Sloan Foundation, the Participating
Institutions, the National Aeronautics and Space Administration, the
National Science Foundation, the U.S. Department of Energy, the
Japanese Monbukagakusho, and the Max Planck Society. The SDSS Web site
is http://www.sdss.org/.  The SDSS is managed by the Astrophysical
Research Consortium (ARC) for the Participating Institutions. The
Participating Institutions are The University of Chicago, Fermilab,
the Institute for Advanced Study, the Japan Participation Group, The
Johns Hopkins University, Los Alamos National Laboratory, the
Max-Planck-Institute for Astronomy (MPIA), the Max-Planck-Institute
for Astrophysics (MPA), New Mexico State University, University of
Pittsburgh, Princeton University, the United States Naval Observatory,
and the University of Washington.  This work was partially supported
by NSF grants AST 99-00703 and AST 03-07582.

\clearpage



\clearpage

\end{document}